\documentclass[runningheads]{llncs}
\usepackage{graphicx}
\usepackage{adjustbox}
\usepackage{caption}
\usepackage{subcaption}
\usepackage{amsmath}
\usepackage{amssymb}
\usepackage{times}
\usepackage{epsfig}
\usepackage{multirow}
\begin{document}
%
\title{RV-GAN: Segmenting Retinal Vascular Structure in Fundus Photographs using a Novel Multi-scale Generative Adversarial Network}
\titlerunning{RV-GAN}
%
\author{Sharif Amit Kamran\inst{1} \and
Khondker Fariha Hossain\inst{1} \and
Alireza Tavakkoli\inst{1} \and 
Stewart Lee Zuckerbrod\inst{2} \and 
Kenton M. Sanders\inst{3} \and 
Salah A. Baker\inst{3}}
\authorrunning{Kamran et al.}
%
\institute{Dept. of Computer Science \& Engineering, University of Nevada, Reno, NV, USA\and Houston Eye Associates, TX, USA\and School of Medicine, University of Nevada, Reno, NV, USA\\ 
}
\maketitle              
\begin{abstract}
High fidelity segmentation of both macro and microvascular structure of the retina plays a pivotal role in determining degenerative retinal diseases, yet it is a difficult problem. Due to successive resolution loss in the encoding phase combined with the inability to recover this lost information in the decoding phase, autoencoding based segmentation approaches are limited in their ability to extract retinal microvascular structure. We propose RV-GAN, a new multi-scale generative architecture for accurate retinal vessel segmentation to alleviate this. The proposed architecture uses two generators and two multi-scale autoencoding discriminators for better microvessel localization and segmentation. In order to avoid the loss of fidelity suffered by traditional GAN-based segmentation systems, we introduce a novel weighted feature matching loss. This new loss incorporates and prioritizes features from the discriminator's decoder over the encoder. Doing so combined with the fact that the discriminator's decoder attempts to determine real or fake images at the pixel level better preserves macro and microvascular structure. By combining reconstruction and weighted feature matching loss, the proposed architecture achieves an area under the curve (AUC)  of  0.9887,  0.9914,  and  0.9887 in pixel-wise segmentation of retinal vasculature from three publicly available datasets,  namely  DRIVE,  CHASE-DB1, and STARE, respectively.  Additionally, RV-GAN outperforms other architectures in two additional relevant metrics, mean intersection-over-union (Mean-IOU) and structural similarity measure (SSIM). 

\end{abstract}
\begin{figure}[t]
    \centering
    \includegraphics[height=9cm,width=0.9\linewidth]{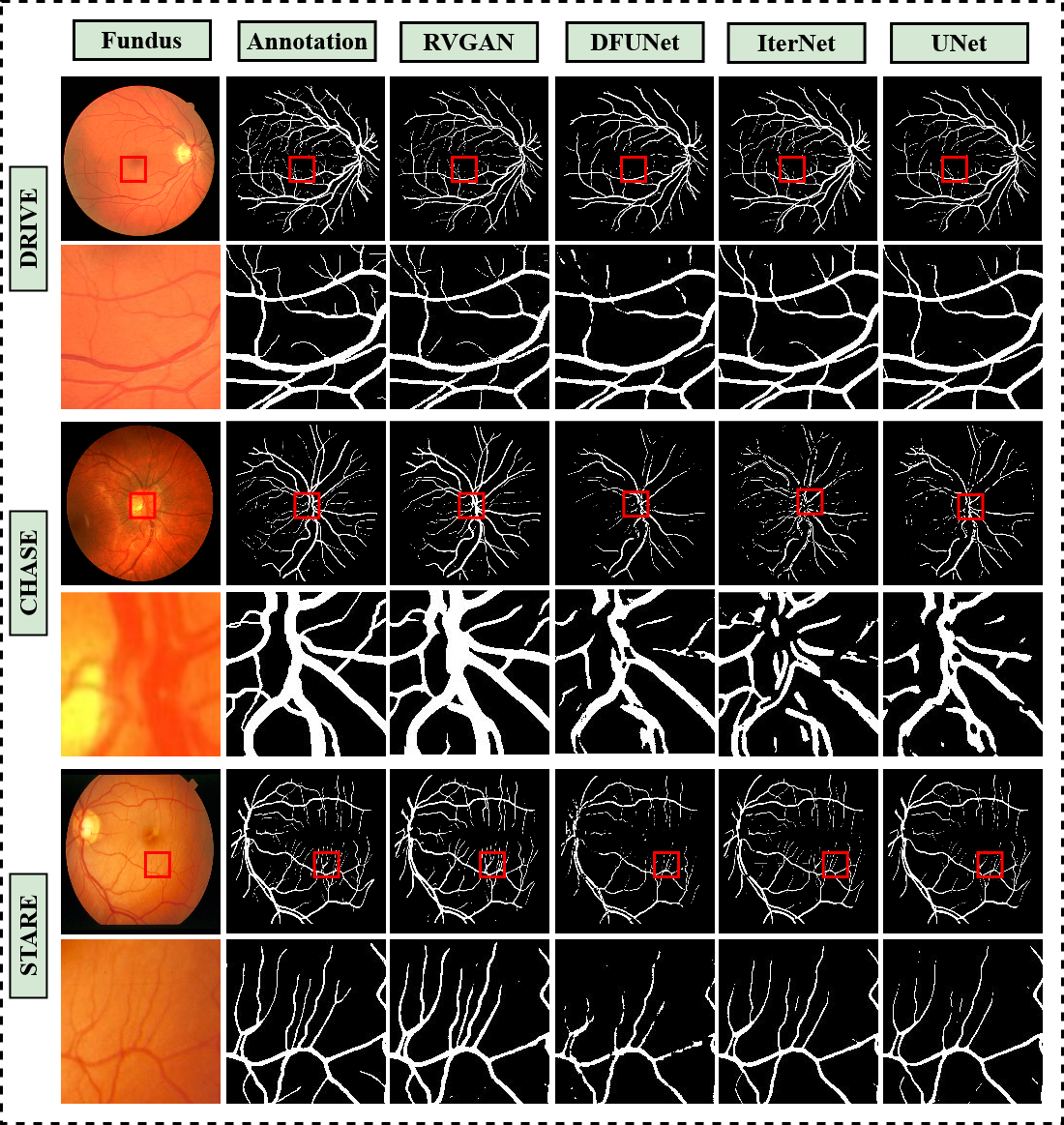}
    \caption{RV-GAN segments vessel with better precision than other architectures. The 1st row is the whole image, while 2nd row is specific zoomed-in area of the image. The Red bounded box signifies the zoomed-in region. Here, the confidence score is, $t>0.5$.  The row contains DRIVE, CHASE-DB1 and STARE data-set. Whereas the column contains fundus images, ground-truths and segmentation maps for RV-GAN, DFUNet, IterNet and UNet. }
    \label{fig1}
\end{figure}
\section{Introduction}
The fundoscopic exam is a procedure that provides necessary information to diagnose different retinal degenerative diseases such as Diabetic Retinopathy, Macular Edema, Cytomegalovirus Retinitis\cite{son2017retinal}. A highly accurate system is required to segment retinal vessels and find abnormalities in the retinal subspace to diagnose these vascular diseases. Many image processing and machine learning-based approaches for retinal vessel segmentation have so far been proposed \cite{soares2006retinal,fraz2012blood,ricci2007retinal,kamran2020improving}. However, such methods fail to precisely pixel-wise segment blood vessels due to insufficient illumination and periodic noises. Attributes like this present in the subspace can create false-positive segmentation \cite{fraz2012blood}. 
In recent times, UNet based deep learning architectures have become very popular for retinal vessel segmentation. UNet consists of an encoder to capture context information and a decoder for enabling precise localization\cite{ronneberger2015u}. Many derivative works based on UNet have been proposed, such as Dense-UNet, Deformable UNet\cite{jin2019dunet}, IterNet \cite{li2020iternet} etc.  These models were able to achieve quite good results for macro vessel segmentation.  However, these architectures fail when segmenting microvessels with higher certainty. One reason is successive resolution loss in the encoder, and failure to capture those features in the decoder results in inferior microvessel segmentation. Recent GAN-based architecture \cite{park2020m,yang2020sud} tries to address this by incorporating discriminative features from adversarial examples while training. However, the discriminator being an encoder\cite{isola2017image}, only trains on patches of images rather than pixels, affecting the true-positive-rate of the model. We need an architecture that can retain discriminative manifold features and segment microvessels on pixel-level with higher confidence. Confidence signifies the probability distribution function of the segmented pixel falling under vessel or background. 
By taking all of these into account, we propose Retinal-Vessel GAN, consisting of coarse and fine generators and multi-scale autoencoder-based discriminators for producing highly accurate segmentation of blood vessel with strong confidence scores. Additionally, we come up with a new weighted feature matching loss with inner and outer weights. And we combine it with reconstruction and hinge loss for adversarial training of our architecture. From Fig.~\ref{fig1}, it is apparent that our architecture produces a segmentation map with a high confidence score. 
\begin{figure}[t]
    \centering
    \includegraphics[height=6cm,width=\linewidth]{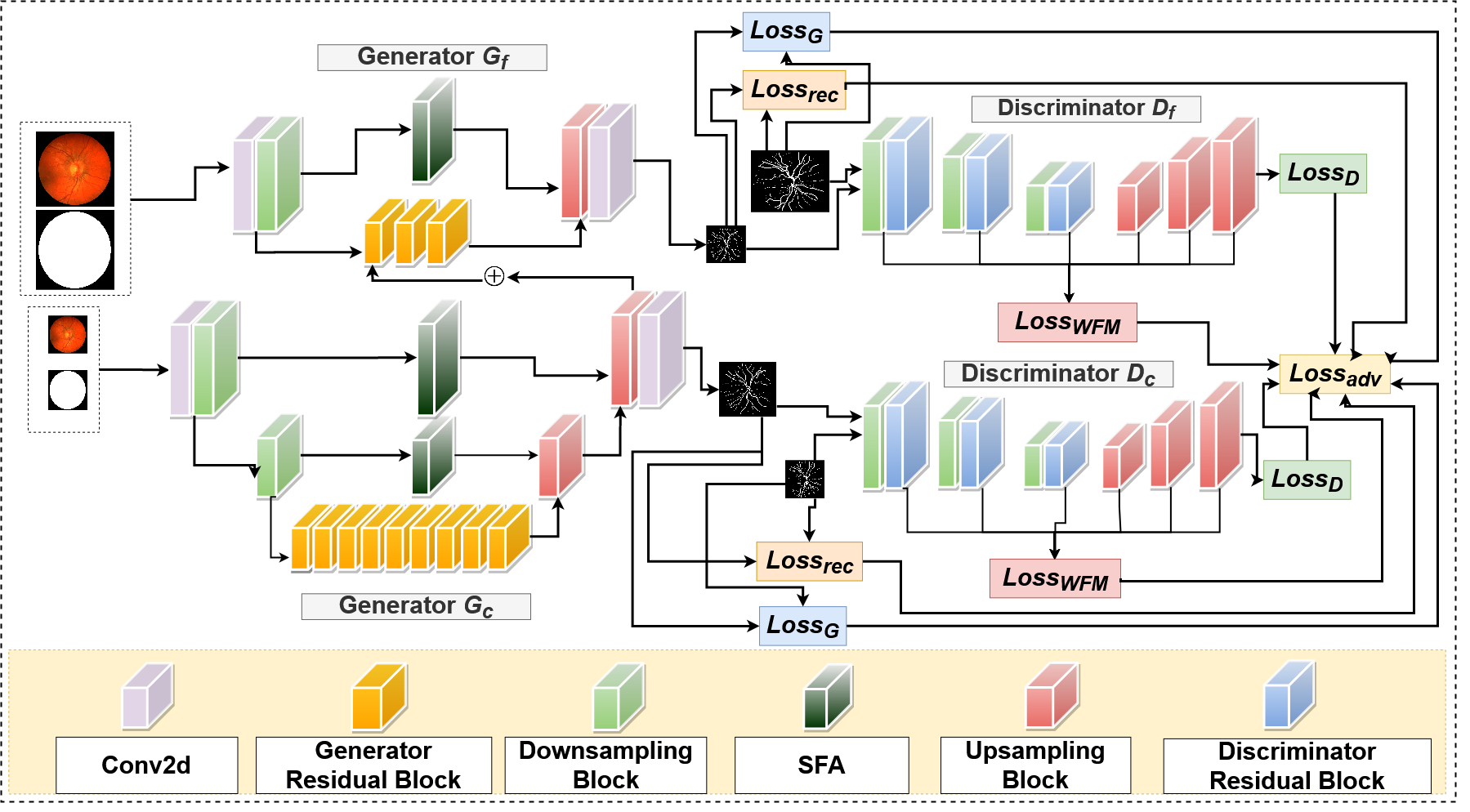}
    \caption{RV-GAN consisting of Coarse and Fine generators $G_f,G_c$ and discriminators $D_f,D_c$. The generators incorporates reconstruction loss, $Loss_{rec}$ and Hinge loss $Loss_{G}$. Whereas the discriminators uses weighted feature matching loss, $Loss_{wfm}$ and Hinge loss $Loss_{D}$. All of these losses are multiplied by weight multiplier and then added in the final adversarial loss, $Loss_{adv}$. The generators consists of Downsampling, Upsampling, SFA and its distinct residual blocks. On the other hand, the discriminators consists of Downsampling, Upsampling and counterpart residual blocks. }
    \label{fig2}
\end{figure}
\section{Proposed Methodology}
\begin{figure}[t]
    \centering
    \includegraphics[width=1\linewidth]{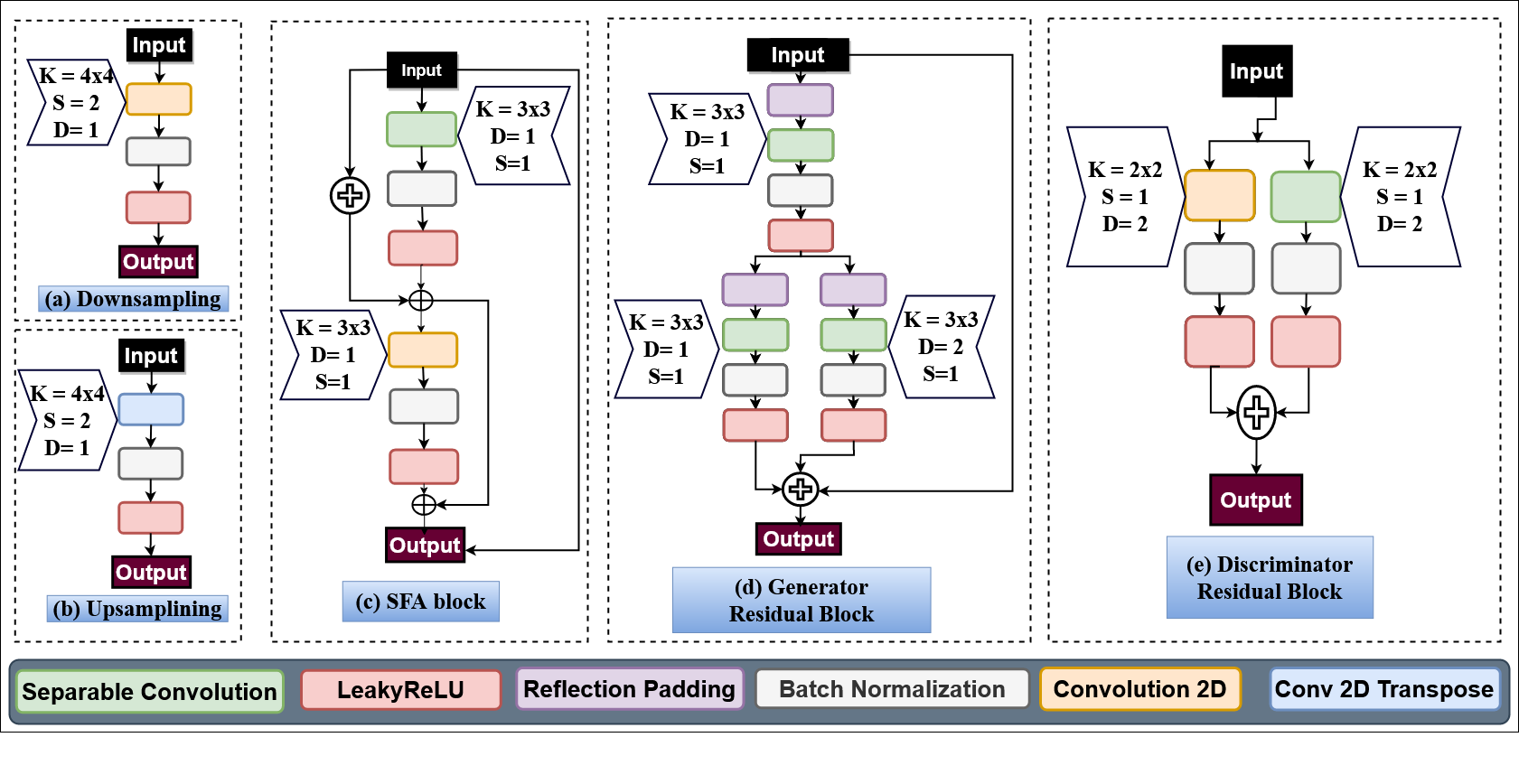}
    \caption{Proposed Downsampling, Upsampling, Spatial Feature Aggregation block, Generator and Discriminator Residual blocks. Here, $K$=Kernel\ size, $S$=Stride, $D$=Dilation. }
    \label{fig3}
\end{figure}
\subsection{Multi-scale Generators}
\label{subsec:generators}
Pairing multi-scale coarse and fine generators produces high-quality domain-specific retinal image synthesis, as observed in recent generative networks, such as Fundus2Angio \cite{kamran2020fundus2angio}, V-GAN \cite{son2017retinal} and \cite{tavakkoli2020novel}. Inspired by this, we also adopt this feature in our architecture by using two generators ($G_{fine}$ and $G_{coarse}$), as visualized in Fig.~\ref{fig2}. The generator $G_{fine}$ synthesizes fine-grained vessel segmentation images by extracting local features such as micro branches, connections, blockages, etc. In contrast, the generator $G_{coarse}$ tries to learn and conserve global information, such as the structures of the maco branches, while producing less detailed microvessel segmentation. The detailed structure of these generators is illustrated in Fig.~\ref{fig2}. 

\subsection{Residual Downsampling and Upsampling Blocks}
\label{subsec:encdec}
Our generators and discriminators consist of both downsampling and upsampling blocks to get the desired feature maps and output. The downsampling block comprises of a convolution layer, a batch-norm layer and a Leaky-ReLU activation function consecutively and is illustrated in Fig.~\ref{fig3}(a). Contrarily, the decoder block consists of a transposed convolution layer, batch-norm, and Leaky-ReLU activation layer successively and can be visualized in  Fig.\ref{fig3}(b). 

\subsection{Distinct Identity Blocks for Generator \& Discriminator}
\label{subsec:residualblock}
For spatial and depth feature propagation, residual identity blocks have become go-to building blocks for image style transfer, image inpainting, and image segmentation tasks \cite{shaham2019singan,wang2018high,choi2020stargan,choi2018stargan,park2019semantic}.Vanilla convolution layers are both computationally inefficient and fail to retain accurate spatial and depth information, as opposed to separable convolution \cite{chollet2017xception}. Separable convolution comprises of a depth-wise convolution and a point-wise convolution successively. As a result, it extracts and preserves depth and spatial features while forward propagating the network. Recent advancement in retinal image classification has shown that combining separable convolutional layers with dilation allows for more robust feature extraction~\cite{opticnet19}. We design two unique residual identity blocks, for our generators and discriminators, as illustrated in Fig.~\ref{fig3}(d) \& Fig.~\ref{fig3}(e).
\subsection{Spatial Feature Aggregation}
\label{subsec:sfa}
In this section, we discuss our proposed spatial feature aggregation (SFA) block, as illustrated in Fig.~\ref{fig3}(c). We use spatial feature aggregation block for combining spatial and depth features from the bottom layers of the network with the top layers of the network, as illustrated in Fig.~\ref{fig2}.   The rationale behind employing the SFA block is to extract and retain spatial and depth information, that is otherwise lost in deep networks. Consequently, these features can be combined with the learned features of the deeper layers of the network to get an accurate approximation, as observed in similar GAN architectures. \cite{zhang2019self,chen2018attention}. 
\subsection{Auto-encoder as Discriminators}
\label{subsec:discriminators}
For better pixel-wise segmentation, we need an architecture that can extract both global and local features from the image. To mitigate this underlying problem, we need a deep and dense architecture with lots of computable parameters. It, in turn, might lead to overfitting or vanishing gradient while training the model. To address this issue, rather than having a single dense segmentation architecture, we adopt light-weight discriminators as autoencoders. Additionally, we use multi-scale discriminators for both our coarse and fine generators, as previously proposed in \cite{li2016precomputed,wang2018high}. The arrangement consists of two discriminators with variable sized input and can help with the overall adversarial training.We define two discriminators as $D_{f}$ and $D_{c}$ as illustrated in Fig.~\ref{fig2}. 

\subsection{Proposed Weighted Feature Matching Loss}
Feature matching loss \cite{wang2018high} was incorporated by extracting features from discriminators to do semantic segmentation. The feature-matching loss is given in Eq.~\ref{eq1} As the authors used Patch-GAN as a discriminator, it only contains an encoding module. By contrast, our work involves finding pixel-wise segmentation of retinal vessel and background and thus requires an additional decoder. By successive downsampling and upsampling, we lose essential spatial information and features; that's why we need to give weightage to different components in the overall architecture. We propose a new weighted feature matching loss, as given in Eq.~\ref{eq2} that combines elements from both encoder and decoder and prioritizes particular features to overcome this. For our case, we experiment and see that giving more weightage to decoder feature map results in better vessel segmentation.
 \begin{equation}
    \mathcal{L}_{fm}(G,D_{enc}) = \mathbb{E}_{x,y} \frac{1}{N}\sum_{i=1}^{k} \Vert D_{enc}^{i}(x,y) - D_{enc}^{i}(x,G(x))\Vert
    \label{eq1}
\end{equation}
\begin{multline}
    \mathcal{L}_{wfm}(G,D_{n}) = \mathbb{E}_{x,y} \frac{1}{N}\sum_{i=1}^{k}\lambda_{enc}^{i} \Vert D_{enc}^{i}(x,y) - D_{enc}^{i}(x,G(x))\Vert + \lambda_{dec}^{i} \Vert D_{dec}^{i}(x,y) \\- D_{dec}^{i}(x,G(x))\Vert
    \label{eq2}
\end{multline}
Eq.~\ref{eq2} is calculated by taking the features from each of the downsampling blocks and upsampling blocks of the discriminator's encoder and decoder. We insert the real and the synthesized segmentation maps consecutively. The N signifies the number of features. Here, $\lambda_{enc}$ and $\lambda_{dec}$ is the inner weight multiplier for each of the extracted feature maps. The weight values are between $[0,1]$, and the total sum of the weight is $1$, and we use a higher weight value for the decoder feature maps than the encoder feature maps.
\subsection{Weighted Objective and Adversarial Loss}
\label{subsec:objective}
For adversarial training, we use Hinge-Loss \cite{zhang2019self,lim2017geometric} as given in Eq.~\ref{eq3} and Eq.~\ref{eq4}. Conclusively, all the fundus images and their corresponding segmentation map pairs are normalized, to $[-1,1]$. As a result, it broadens the difference between the pixel intensities of the real and synthesized segmentation maps. In Eq.~\ref{eq5}, we multiply $\mathcal{L}_{adv}(G)$ with $\lambda_{adv}$ as weight multiplier. Next, we add $\mathcal{L}_{adv}(D)$ with the output of the multiplication.
\begin{equation}
    \mathcal{L}_{adv}(D) = - \mathbb{E}_{x,y} \big[\ \min(0,-1+D(x,y))\big]\ \\-  \mathbb{E}_{x} \big[\ \min(0,-1-D(x,G(x))) \big]\
    \label{eq3}
\end{equation}
\begin{equation}
    \mathcal{L}_{adv}(G) = - \mathbb{E}_{x,y} \big[(D(G(x),y))\big]\
    \label{eq4}
\end{equation}
\begin{equation}
    \mathcal{L}_{adv}(G,D) = \mathcal{L}_{adv}(D) + \lambda_{adv} (\mathcal{L}_{adv}(G)) 
    \label{eq5}
\end{equation}
In Eq.~\ref{eq4} and Eq.~\ref{eq5}, we first train the discriminators on the real fundus, $x$ and real segmentation map, $y$. After that, we train with the real fundus, $x$, and synthesized segmentation map, $G(x)$. We begin by batch-wise training the discriminators $D_{f}$, and $D_{c}$ for a couple of iterations on the training data. Following that, we train the $G_{c}$ while keeping the weights of the discriminators frozen. In a similar fashion, we train $G_{f}$ on a batch training image while keeping weights of all the discriminators frozen. 

The generators also incorporate the reconstruction loss (Mean Squared Error) as shown in Eq.~\ref{eq6}. By utilizing the loss we ensure the synthesized images contain more realistic microvessel, arteries, and vascular structure.
\begin{equation}
    \mathcal{L}_{rec}(G) = \mathbb{E}_{x,y} \Vert G(x) - y \Vert^2
    \label{eq6}
\end{equation}

By incorporating Eq.~\ref{eq2}, \ref{eq5} and \ref{eq6}, we can formulate our final objective function as given in Eq.~\ref{eq7}.
\begin{multline}
\min \limits_{G_{f},G_{c}} \big( \max \limits_{D_{f},D_{c}}  (\mathcal{L}_{adv}(G_{f},G_{c}, D_{f},D_{c})) + \\ \lambda_{rec}\big[\ \mathcal{L}_{rec}(G_{f},G_{c})\big]\ + 
\lambda_{wfm}\big[\ \mathcal{L}_{wfm}(G_{f},G_{c}, D_{f},D_{c})\big]\ \big)
\label{eq7}
\end{multline}
Here, $\lambda_{adv}$, $\lambda_{rec}$, and $\lambda_{wfm}$ implies different weights, that is multiplied with their respective losses. The loss weighting decides which architecture to prioritize while training. For our system, more weights are given to the $\mathcal{L}_{adv}(G)$, $\mathcal{L}_{rec}$, $\mathcal{L}_{wfm}$, and thus we select bigger $\lambda$ values for those. 
\begin{figure}[t]
    \centering
    \begin{subfigure}[]{0.3\textwidth}
        \centering
        \includegraphics[height=1.2in]{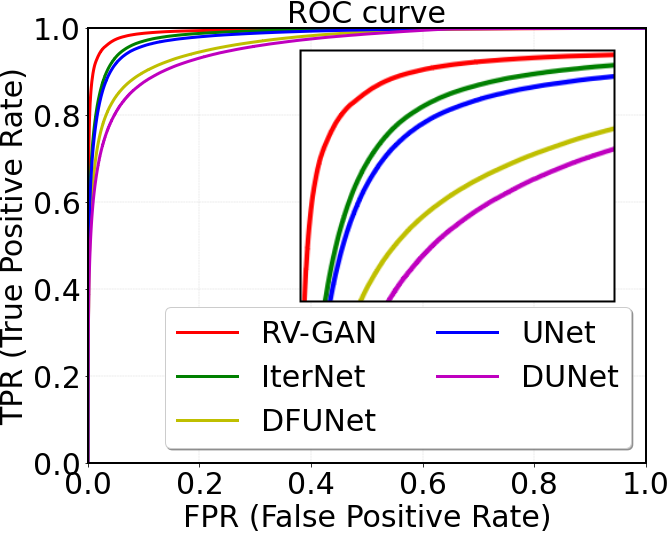}
        \caption{DRIVE}
    \end{subfigure}%
    ~ 
    \begin{subfigure}[]{0.3\textwidth}
        \centering
        \includegraphics[height=1.2in]{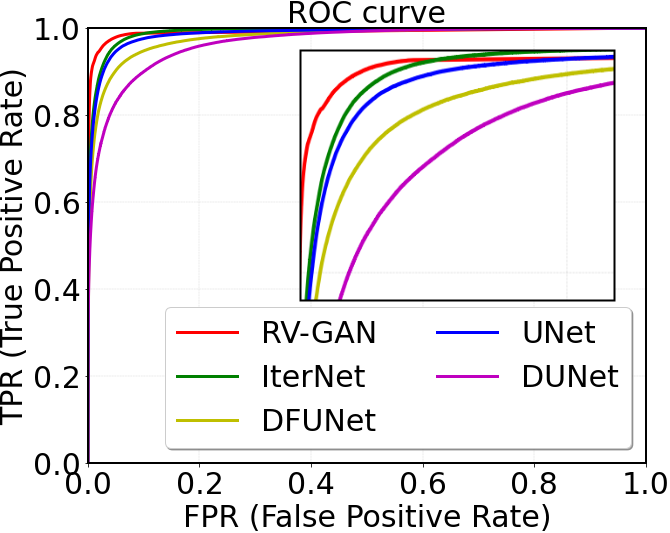}
        \caption{STARE}
    \end{subfigure}
    ~ 
    \begin{subfigure}[]{0.3\textwidth}
        \centering
        \includegraphics[height=1.2in]{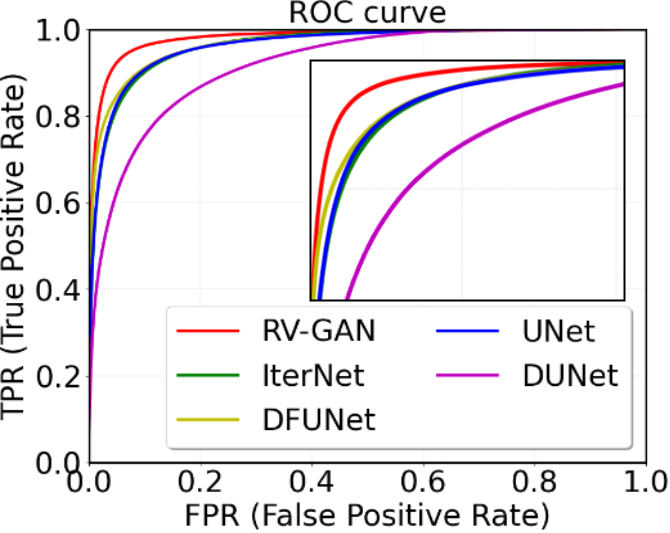}
        \caption{CHASE-DB1}
    \end{subfigure}
    \caption{ROC Curves on (a) DRIVE (b) STARE (c) CHASE-DB1.}
    \label{fig5}
\end{figure}

\section{Experiments}
\subsection{Dataset}
\label{subsec:dataset}
For benchmarking, we use three retinal segmentation datasets, namely,  DRIVE  \cite{staal2004ridge}, CHASE-DB1 \cite{owen2009measuring},  and  STARE \cite{hoover2000locating}.  The images are respectively in $.tif(565\times584)$, $.jpg(999\times960)$, and $.ppm(700\times605)$. We train three different RV-GAN networks with each of these datasets using 5-fold cross-validation.  We use overlapping image patches with a stride of 32 and an image size of $128\times128$ for training and validation. So we end up having 4320 for STARE,  15120 for CHASE-DB1, and 4200 for DRIVE from 20, 20, and 16 images.  DRIVE dataset comes with official FoV masks for the test images.  For CHASE-DB1 and STARE dataset, we also generate FoV masks similar to Li et al \cite{li2020iternet}.  For testing,  overlapping image patches with a stride of 3 were extracted and averaged by taking 20, 8 and 4 images from DRIVE, CHASE-DB1, and STARE.

\subsection{Hyper-parameter Initialization}
\label{subsec:hyper}
For adversarial training, we used hinge loss \cite{zhang2019self,lim2017geometric}. We picked $\lambda_{enc}=0.4$ (Eq.~\ref{eq1}), $\lambda_{dec}=0.6$ (Eq.~\ref{eq2}), $\lambda_{adv}=10$ (Eq.~\ref{eq5}), $ \lambda_{rec} =10$ (Eq.~\ref{eq6}) and $ \lambda_{wfm} =10$ (Eq.~\ref{eq7}). We used Adam optimizer \cite{kingma2014adam}, with learning rate $\alpha=0.0002$, $\beta_1=0.5$ and $\beta_2=0.999$. We train with mini-batches with batch size, $b=24$ for 100 epochs in three stages using Tensorflow. It took between 24-48 hours to train our model on NVIDIA P100 GPU depending on data-set. Because DRIVE and STARE have lower number of patches compared to CHASE-DB1, it takes less amount to train. The inference time is $0.025$ second per image.
\subsection{Threshold vs. Confidence Score}
Confidence score signifies the per-pixel probability density value of the segmentation map. U-Net-derived models incorporate binary cross-entropy loss with a threshold of $0.5$ to predict if the pixel is a vessel or background. So a pixel, predicted with a $0.5001$ probability, will be classified as a vessel. As a result, the model suffers from Type I error or a high false-positive rate (FPR). In contrast, we use the generators to produce realistic segmentation maps and utilize the weighted feature matching loss to combine inherent manifold information to predict real and fake pixels with higher certainty. Consequently, we can see in Fig.~\ref{fig5} that our model's Receiver Operating (ROC) curves for three data-sets are relatively better than other previous methods due to a high confidence score.

\begin{table}[t!]
\centering
\caption{Performance comparison on DRIVE \cite{staal2004ridge},  CHASE-DB1 \cite{owen2009measuring}, \& STARE \cite{hoover2000locating}.}
\centering
\begin{adjustbox}{width=1\textwidth}
\begin{tabular}{c|c|c|c|c|c|c|c|c|c}
\hline
Dataset & Method & Year & F1 Score & Sensitivity & Specificity & Accuracy & AUC-ROC & Mean-IOU & SSIM \\ \hline
\multirow{9}{*}{DRIVE} & UNet \cite{jin2019dunet} & 2018 &  0.8174 &   0.7822 &  0.9808 & 0.9555 &  0.9752 & 0.9635 & 0.8868 \\ 
& Residual UNet \cite{alom2018recurrent} & 2018 & 0.8149 & 0.7726 & 0.9820 & 0.9553 & 0.9779 & -  & - \\ 
&Recurrent UNet \cite{alom2018recurrent} & 2018 & 0.8155 & 0.7751 & 0.9816 & 0.9556 & 0.9782  & - & -\\ 
& R2UNet \cite{alom2018recurrent} & 2018 & 0.8171 &0.7792 & 0.9813 & 0.9556 & 0.9784 & - &  -     \\ 
 & DFUNet \cite{jin2019dunet} & 2019 &  0.8190 & 0.7863 &  0.9805 &   0.9558 &  0.9778  & 0.9605 & 0.8789     \\
& IterNet \cite{li2020iternet} & 2019 & 0.8205 &  0.7735 & 0.9838 & 0.9573 & 0.9816& 0.9692 & 0.9008\\
& SUD-GAN \cite{yang2020sud}& 2020 & - & 0.8340 & 0.9820 & 0.9560 & 0.9786 & - & - \\
& M-GAN \cite{park2020m} & 2020 & 0.8324 & \textbf{0.8346} & 0.9836 & 0.9706 & 0.9868 & - & - \\
& \textbf{RV-GAN (Ours)} & 2020 & \textbf{0.8690} & 0.7927 & \textbf{0.9969} & \textbf{0.9790} & \textbf{0.9887} & \textbf{0.9762} & \textbf{0.9237}  \\\hline	
\multirow{6}{*}{CHASE-DB1}& UNet \cite{jin2019dunet} & 2018 & 0.7993 & 0.7841 & 0.9823 & 0.9643 & 0.9812 & 0.9536 & 0.9029\\
& DenseBlock-UNet \cite{li2018h}  & 2018 & 0.8006 & 0.8178 & 0.9775 & 0.9631 & 0.9826 &0.9454 & 0.8867 \\ 
& DFUNet \cite{jin2019dunet} & 2019 & 0.8001 & 0.7859 & 0.9822 & 0.9644 & 0.9834 &  0.9609 & 0.9175     \\
& IterNet \cite{li2020iternet} & 2019 & 0.8073 & 0.7970 & 0.9823 & 0.9655 & 0.9851 & 0.9584 & 0.9123\\
& M-GAN \cite{park2020m} & 2020 & 0.8110 & 0.8234 & \textbf{0.9938} & \textbf{0.9736} & 0.9859 & - & - \\
& \textbf{RV-GAN (Ours)} & 2020 & \textbf{0.8957} & \textbf{0.8199} & 0.9806 & 0.9697 & \textbf{0.9914} & \textbf{0.9705} & \textbf{0.9266}  \\\hline
\multirow{7}{*}{STARE}& UNet \cite{jin2019dunet} & 2018 & 0.7595 & 0.6681 & 0.9915 & 0.9639 & 0.9710 & 0.9744 & 0.9271\\
& DenseBlock-UNet \cite{li2018h}  & 2018  & 0.7691 & 0.6807 & 0.9916 & 0.9651 & 0.9755 & 0.9604 & 0.9034 \\ 
& DFUNet \cite{jin2019dunet} & 2019 & 0.7629 & 0.6810 & 0.9903 & 0.9639 & 0.9758 & 0.9701 &  0.9169      \\
& IterNet \cite{li2020iternet} & 2019 & 0.8146 & 0.7715 & 0.9886 & 0.9701 & 0.9881 & 0.9752 & 0.9219\\
& SUD-GAN \cite{yang2020sud}& 2020 & - & 0.8334 & 0.9897 & 0.9663 & 0.9734 & - & - \\
& M-GAN \cite{park2020m} & 2020 & \textbf{0.8370} & 0.8234 & \textbf{0.9938} & \textbf{0.9876} & 0.9873 & - & - \\
& \textbf{RV-GAN (Ours)} & 2020 & 0.8323 & \textbf{0.8356} & 0.9864 & 0.9754 & \textbf{0.9887} & \textbf{0.9754} & \textbf{0.9292}  \\\hline	
\end{tabular}
\label{table1}
\end{adjustbox}
\end{table}

\subsection{Quantitative Bench-marking}
\label{subsec:quant}
We compared our architecture with some best performing ones, including UNet \cite{jin2019dunet}, DenseBlock-UNet \cite{li2018h}, Deform-UNet \cite{jin2019dunet} and IterNet \cite{li2020iternet} as illustrated in Fig.~\ref{fig1}. We trained and evaluated the first three architectures using their publicly available source code by ourselves on the three datasets. For IterNet, the pre-trained weight was provided, so we used that to get the inference result. Next, we do a comparative analysis with existing retinal vessel segmentation architectures, which includes both UNet and GAN based models. The prediction results for DRIVE, CHASE-DB1 and STARE are provided in Table.~\ref{table1}. We report traditional metrics such as  F1-score, Sensitivity, Specificity, Accuracy, and AUC-ROC. Additionally, we use two other metrics for predicting accurate segmentation and structural similarity of the retinal vessels, namely Mean-IOU (Jaccard Similarity Coefficient) and Structural Similarity Index\cite{wang2004image}. We chose Mean-IOU because its the gold standard for measuring segmentation results for many Semantic Segmentation Challenges such as Pascal-VOC2012 \cite{everingham2015pascal}, MS-COCO \cite{lin2014microsoft}. Contrarily, SSIM is a standard metric for evaluating GANs for image-to-image translation tasks. As illustrated in all the tables, our model outperforms both UNet derived architectures and recent GAN based models in terms of AUC-ROC, Mean-IOU, and SSIM, the three main metrics for this task. M-GAN achieves better Specificity and Accuracy in CHASE-DB1 and STARE. However, higher Specificity means better background pixel segmentation (True Negative), which is less essential than having better retinal vessel segmentation (True Positive).  We want both, better Sensitivity and AUC-ROC, which equates to having a higher confidence score. In Fig.~\ref{fig5} we can see that our True positive Rate is always better than other architectures for all three data-set. We couldn't report SSIM and Mean-IOU for some of the architectures as source codes and pre-trained, weights weren't provided.

\section{Conclusion}
In this paper,  we proposed a  new multi-scale generative architecture called RV-GAN. By combining our novel featuring matching loss, the architecture synthesizes precise venular structure segmentation with high confidence scores for two relevant metrics.  As a result,  we can efficiently employ this architecture in various applications of ophthalmology. The model is best suited for analyzing retinal degenerative diseases and monitoring future prognosis. We hope to extend this work to other data modalities.

 \bibliographystyle{splncs04}
\bibliography{reference}
\end{document}